\documentclass[aps,showpacs]{revtex4-2}
\usepackage{graphicx}
\usepackage{bm}
\usepackage{graphicx}
\usepackage{amsmath}
\usepackage{mathrsfs}
\usepackage{ulem}
\usepackage{color}
\usepackage[colorlinks, linkcolor=red,anchorcolor=green,citecolor=blue]{hyperref}

\newcommand{\nn}{\nonumber}
\newcommand\diag{\operatorname{diag}}

\graphicspath{{./figs/}}

\begin{document}  
	
\title{Relativistic first-order spin hydrodynamics via the Chapman-Enskog expansion }

\author{Jin Hu}
\email{hu-j17@mails.tsinghua.edu.cn}
\affiliation{Department of Physics, Tsinghua University, Beijing 100084, China}

\begin{abstract}
In this paper, we present a detailed derivation of relativistic first-order spin hydrodynamics using the Chapman-Enskog method to linearize the nonlocal collision term for massive fermions proposed in \cite{Weickgenannt:2021cuo}, which well  describes  spin-orbit coupling in the collision process and is relevant for the research on local spin polarization. Based on this collisional term, we provide a formal discussion about first-order spin hydrodynamics and determine the motion equations of fluid variables and nonequilibrium corrections to the energy-momentum and spin tensors. The results indicate that  the motion equations show no differences compared to spinless first-order hydrodynamics and the energy-momentum tensor receives no corrections from spin as far as  first-order theory is concerned, which calls for the construction of second-order theory of fluids naturally incorporating the effect of spin-orbit coupling.
\end{abstract}




\maketitle	
\section{Introduction}
Recent developments in the experiments of relativistic non-central heavy-ion collisions have seen great progress in measuring the spin-related observables of $ \Lambda $ hyperons, the results of which show the phenomena of  spin polarization \cite{STAR:2017ckg,Alpatov:2020iev}. This originates from the conversion of orbital angular momentum carried by the media, quark-gluon plasma (QGP),  into spin angular momentum via the well-known mechanism of orbit-spin coupling.  Many theoretical researches  on the global polarization of $ \Lambda $ hyperons have been carried out and provide a good description of experimental data. See related works in \cite{Wei:2018zfb,Karpenko:2016jyx,Csernai:2018yok,Li:2017slc,Bzdak:2017shg,Shi:2017wpk,Sun:2017xhx,Ivanov:2019wzg,Xie:2017upb}.  Afterward, the STAR Collaboration published   measurements of differential spin polarization, i.e. the dependance of $ \Lambda $ polarization  on the azimuthal angle and transverse momentum \cite{Adam:2019srw,Adam:2018ivw}. However, theoretical calculation following previous models can not explain the experimental data satisfyingly and even predicts the opposite dependance, which is usually called ``spin sign problem'' \cite{Becattini:2017gcx,Xia:2018tes}. Resolving this problem calls for new theoretical frameworks. Among all the candidates, spin hydrodynamics is a promising one which extends ordinary fluid models by including the spin degree of freedom. On the other hand, these direct experimental measurements of quantum effects in relativistic heavy-ion collisions provide the opportunity to study  the evolution of spinful fluids.

There are many efforts following this direction to try to get insight into the spin sign prolem. ``Ideal'' spin hydrodynamics was proposed in the context of the QGP~\cite{Florkowski:2017ruc} and  for massive spin-$1/2$ fermions \cite{Peng:2021ago}. See also relevant discussions in \cite{Becattini:2009wh, Florkowski:2018fap}. Recently, first-order spin hydrodynamics including non-equilibrium corrections has also been put into efforts~\cite{Hattori:2019lfp,Fukushima:2020ucl,Hu:2021lnx} based on general arguments. The related research work on first-order spin hydrodynamics can be also found in \cite{Bhadury:2020cop} using the method of relaxation time approximation (RTA). Though with great progress in investigating spinful fluids, it is noted that the ultimate goal for us is to construct a causal and numerically stable theory of spin hydrodynamics, which allows the numerical implementation and simulation of the evolution of the fluid system, and eventually provides us with quantitative explainations of the spin-related experimental phenomena. To that end, the hydrodynamic equations and relevant transport coefficients must be obtained in a first-principle fashion, which equivalently means that the macroscopic evolution of spinful fluids is dictated by the microscopic theory of spin transport. Therefore it is necessary to derive the quantum transport equations with proper collision terms, which lays the basis for the construction of a causal and numerically stable theory of spin hydrodynamics. The related developments in spin transport can be found in \cite{Weickgenannt:2021cuo,Weickgenannt:2020aaf,Yang:2020hri,Wang:2020pej,Sheng:2021kfc}.

In this paper, we try to derive relativistic first-order spin hydrodynamics from microscopic theory, which is based on the framework of the Chapman-Enskog expansion. To make it,  the transport equation along with the nonlocal collision term proposed in \cite{Weickgenannt:2021cuo} is adopted as the startpoint of our derivation.   
This paper is organized as follows. In Sec.~\ref{secqu}
we present a brief review of the relativistic transport equation with the nonlocal collision term. After that, the local equilibrium distribution is discussed and the conditions for global equilibrium are determined in Sec.~\ref{eq}. In Sec.~\ref{otion}, we derive the motion equations of all hydrodynamic variables. In Sec.~\ref{expansion} we adopt the  Chapman-Enskog expansion to derive  first-order spin hydrodynamics and provide the detailed form of the deviation function from local equilibrium distribution function up to first order in gradients. The corrections to various tensors we are concerned about are calculated in \ref{corr}. Discussions and outlook are given in Sec.~\ref{su}. Natural units $k_B=c=\hbar=1$ are used. The metric tensor here is given by $g^{\mu\nu}=\diag(1,-1,-1,-1)$ , while $\Delta^{\mu\nu} \equiv g^{\mu\nu}-u^\mu u^\nu$ is the projection tensor orthogonal to the four-vector fluid velocity $u^\mu$.

In addition, we employ the symmetric/antisymmetric shorthand notations:
\begin{eqnarray}
X^{( \mu\nu ) } &\equiv& (X^{ \mu\nu } + X^{ \nu \mu})/2, \\
X^{[ \mu\nu ] } &\equiv& (X^{ \mu\nu } - X^{ \nu \mu})/2, \\
X^{\langle \mu\nu \rangle}&\equiv&
	\bigg(\frac{\Delta^{\mu}_{\alpha} \Delta^{\nu}_{\beta} 
		+ \Delta^{\nu}_{\alpha} \Delta^{\mu}_{\beta}}{2}
	- \frac{\Delta^{\mu\nu} \Delta_{\alpha\beta}}{3}\bigg)X^{\alpha\beta}.
\end{eqnarray}
Specially, we decompose the derivative $\partial$ according to 
\begin{align}
\partial^\mu=u^\mu D+\nabla^\mu,\quad D\equiv u^\mu\partial_\mu,\quad \nabla^\mu\equiv\Delta^{\mu\nu}\partial_\nu.
\end{align}
And the following notations for the derivatives of $u^\mu$ are often used.
\begin{align}
\sigma^{\mu\nu}\equiv\nabla^{\langle\mu}u^{\nu\rangle}, \quad \theta\equiv\nabla_\mu u^\mu.
\end{align}

\section{nonlocal transport equations and extented phase space}
\label{secqu}	
We start with a quantum	transport equation with a nonlocal collision term for massive fermions 
which is presented in \cite{Weickgenannt:2021cuo}.
The evolution of the system is governed by the proposed on-shell Boltzmann equation incorporating the variable $\bm{s}$ as a classical description of spin degrees of freedom,
\begin{align}
\label{boltz}
\delta(p^2-m^2)p\cdot \partial f(x,p,\bm{s})=\delta(p^2-m^2)C[f],
\end{align}	 
with the nonlocal collision term given by
\begin{eqnarray}
\label{cf}
C[f] &  \equiv& \int d\Gamma_1 d\Gamma_2 d\Gamma^\prime\,    
\mathcal{W}\,  
[f(x+\Delta_1,p_1,\bm{s}_1)f(x+\Delta_2,p_2,\bm{s}_2)
-f(x+\Delta,p,\bm{s})f(x+\Delta^\prime,p^\prime,\bm{s}^\prime)]\nn\\
&+& \int  d\Gamma_2 \, dS_1(p)\,\mathcal{W}_2f(x+\Delta_1,p,\bm{s}_1)
f(x+\Delta_2,p_2,\bm{s}_2)\;, \label{finalcollisionterm} 
\end{eqnarray}
with the measure defined as $d\Gamma \equiv d^4p\, \delta(p^2 - m^2)dS(p)$ and the newly introduced measure $dS(p)$ reflects the extended phase space now include spin, which will be further discussed hereafter.
 Here the transition rates $\mathcal{W}$ is  shown by
\begin{eqnarray}
\mathcal{W}&\equiv& \delta^{(4)}(p+p^\prime-p_1-p_2)\,
\frac{1}{8} \sum_{s,r}   h_{s r} (p,\bm{s})  
\sum_{s',r',s_1,s_2,r_1,r_2} h_{s^\prime r^\prime}(p^\prime, \bm{s}^\prime) \,  
h_{s_1 r_1}(p_1, \bm{s}_1)\, h_{s_2 r_2}(p_2, \bm{s}_2) \nn \\
&&\times \langle{p,p^\prime;r,r^\prime|t|p_1,p_2;s_1,s_2}\rangle
\langle{p_1,p_2;r_1,r_2|t^\dagger|p,p^\prime;s,s^\prime}\rangle \;
\label{local_col_GLW_after}
\end{eqnarray}
with 
\begin{equation}
h_{s r}(p,\bm{s}) \equiv \delta_{s r}+  \frac{1}{2m}\, \bar{u}_s(p)\gamma^5\bm{s} \cdot\gamma u_r(p)\; ,
\end{equation} 
where the $\gamma$ matrices, spinor $u_s(p)$ and spin indices $r,s$ above all correspond to the spinor description for fermions as often used, and the matrix element of $t$ is the conventional scattering amplitude defined in quantum field theory.
To proceed, we briefly review the transport equation with the collision term. The crucial point for the nontrivial extension of the collision term is reflected in the Enskog-type shift $\Delta$ manifesting the nonlocality of collisions with its definition given by
\begin{equation}
\label{deltanon}
\Delta^\mu\equiv -\frac{1}{2m(p\cdot\hat{t}+m)}\, 
\epsilon^{\mu\nu\alpha\beta}p_\nu \hat{t}_\alpha \bm{s}_{\beta}\;,
\end{equation}
where $\hat{t}^\mu$ is the time-like unit vector which is $(1,\boldsymbol{0})$ 
in the frame where $p^\mu$ is measured.
To note, such an appealing structure of the transport equation  originates from the nontrivial tensor structure of particle fields, or equivalently the nontrivial dynamics introduced by spin angular momentum compared to related discussions about scalar field in \cite{Hu:2021plu}. A conclusion can be drawn that this shift well captures the properties of spin-orbit coupling in nonlocal collisions, which is highly relevant for solving the spin sign problem of local polarization of $\Lambda$. 
Here we only focus on the first term  Eq.(\ref{cf}) which
describes momentum- and spin-exchange interactions,
while the second term  {}{corresponds to} 
spin exchange without momentum exchange \cite{DeGroot:1980dk}.
When neglecting spin, $h_{sr}\rightarrow \delta_{sr}$ and the collision term recovers the widely used local form of two-body scattering. 
 
 It is now time to fix our focus on the classical description of spin. Spin here is treated as an additional variable in phase space
 \cite{Zamanian:2010zz,Ekman:2017kxi,Ekman:2019vrv,Florkowski:2018fap,Bhadury:2020puc,Weickgenannt:2020aaf}, 
 which is  immediately connected to  the first-principle 
 quantum description to a ``classical'' description of spin. 
 Moreover, the authors  of \cite{Weickgenannt:2021cuo} make good use of this concept to combine the full dynamics of the Wigner function into one scalar equation and gives a natural 
 interpretation for the conservation laws 
 and the collisional invariants. 
  Then the covariant
  integration measure for spin is presented as 
  \begin{equation}
  \int dS(p)  \equiv \sqrt{\frac{p^2}{3 \pi^2}} \int d^4\bm{s}\, \delta(\bm{s}\cdot\bm{s}+3)
  \delta(p\cdot \bm{s})\;,
  \end{equation}
 And the following properties are helpful for our later calculation.
  \begin{subequations}
  	\begin{eqnarray}
  	\label{f1}
  	\int dS(p) & = & 2\;, \\
  	\label{f2}
  	\int dS(p)\, \bm{s}^\mu & = & 0 \;, \\
  	\label{f3}
  	\int dS(p)\, \bm{s}^\mu \bm{s}^\nu & = & - 2 \left( g^{\mu \nu} - \frac{p^\mu p^\nu}{p^2} \right)\;,\\
  	\label{f31}
  	\int dS(p)\, \bm{s}^\mu \bm{s}^\nu \bm{s}^\rho& = & 0\;,\\  	
  	\label{f4}
\int dS(p) \Sigma^{\mu\nu}_{\bm{s}}\Sigma_{\bm{s}}^{\alpha\beta}
&=&\frac{2}{m^2}(g^{\mu\alpha}g^{\nu\beta}p^2+g^{\mu\beta}p^{\nu}p^{\alpha}+g^{\nu\alpha}p^{\mu}p^{\beta}-[\mu\leftrightarrow\nu]).  	 
  	\label{dS_smu_snu}
  	\end{eqnarray}
  \end{subequations}
  Therefore the tensors particle current, energy-momentum tensor and spin tensor we are concerned about can be written as
 \begin{align}
 \label{N}
 &N^{\mu}\equiv\int\,d\Gamma \,p^\mu f(x,p,\bm{s}),\\
 \label{T}
 &T_{\text{HW}}^{\mu\nu}\equiv\int \,d\Gamma \,p^\mu p^\nu f(x,p,\bm{s}),\\
 \label{S}
 &S_{\text{HW}}^{\lambda,\mu\nu}\equiv\int \,d\Gamma \,p^\lambda (\frac{1}{2}\Sigma^{\mu\nu}_{\bm{s}}-\frac{1}{2m^2}p^{[\mu}\partial^{\nu]} )f(x,p,\bm{s}),
 \end{align}
 where we have chosen the psudo-gauge proposed by Hilgevoord and Wouthuysen (HW) \cite{HILGEVOORD19631,hilgevoord1965covariant}. When including non-local collisions, $T_{\text{HW}}^{\mu\nu}$ has anti-symmetric component belonging to the order of $O(\partial^2)$ \cite{Weickgenannt:2020aaf}, which is neglected in our constructing first-order theory.
 In the following sections, when nothing confusing occurs, the subscript HW will be omited.
 
  \section{Equilibrium}
  \label{eq}	
  In this section, we will show that the collision term Eq.(\ref{cf}) is consistent with the standard form of spin-dependent local equilibrium distribution function \cite{Becattini:2013fla,Florkowski:2017ruc},
  \begin{align}
  \label{feq}
  &f_{\text{leq}}(x,p,\bm{s})=\frac{1}{(2\pi)^3}\exp[\xi-\beta\cdot p+\frac{\Omega_{\mu\nu}\Sigma_{\bm{s}}^{\mu\nu}}{4}], 
  \end{align}
  where $\Omega_{\mu\nu}$ represents spin potential, while $\beta^\mu\equiv\frac{u^\mu}{T},\xi\equiv \frac{\mu}{T},\Sigma_{\bm{s}}^{\mu\nu}\equiv -\frac1m \epsilon^{\mu\nu\alpha\beta}p_\alpha \bm{s}_\beta$ with the temperature $T$,  and  the chemical potential $\mu$  introduced for conserved particle number (only elastic scatterings are considered). The exponent in Eq.(\ref{feq}) is exactly the linear combination of all conserved quantities, and $\xi,\beta$ and $\Omega^{\mu\nu}$ are the correspondent Lagrangian multipliers maximizing the total entropy of the system. To prove this, the substitution of Eq.(\ref{feq}) into  Eq.(\ref{cf}) leads to
  \begin{eqnarray}
  \label{cf00}
  C[f_{\text{leq}}] 
  &= & -\frac{1}{(2\pi)^6} \int d\Gamma^\prime d\Gamma_1 d\Gamma_2  \,
  \mathcal{W} \exp(2\xi-\beta\cdot(p+p^\prime))\nn\\
  & &\times  \big[-\partial_\mu\xi \frac{}{}
  \left(\Delta_1^\mu +\Delta_2^\mu -\Delta^\mu -\Delta^{\prime\mu}
  \right)+\partial_\mu\beta_\nu \frac{}{}
  \left(\Delta_1^\mu p_1^\nu+\Delta_2^\mu p_2^\nu-\Delta^\mu p^\nu-\Delta^{\prime\mu}
  p^{\prime\nu} \right)\nn\\
  &&
  \quad\,\,- \frac 1 4\Omega_{\mu\nu}\left(\Sigma_{\bm{s}_1}^{\mu\nu}
  +\Sigma_{\bm{s}_2}^{\mu\nu}
  -\Sigma_{\bm{s}}^{\mu\nu}-\Sigma_{\bm{s}^\prime}^{\mu\nu}\right) \big],
  \label{colleqq}
  \end{eqnarray}
  where the local equilibrium distribution is taylor expanded to first order in $\Omega$ assuming small spin potential (\,if the system in discussion is close to the state of global equilibrium, $\Omega$ is about the order of the gradient of $\beta$ field\,). 
  Assuming total angular momentum $J^{\mu\nu}=2\Delta^{[\mu}p^{\nu]}+\frac{1}{2}\Sigma^{\mu\nu}_{\bm{s}}$ is conserved in collisions, we can conclude that  it is the global equilibrium distribution function that makes the collision term vanish contrary to traditional definition for local equilibrium when including spin. In that case, the conditions for vanishing Eq.(\ref{colleqq}) are
  \begin{align}
  \label{condition}
  &\partial_{(\mu}\beta_{\nu)}=0,\quad \xi=\text{const},\nn\\
  &\Omega_{\mu\nu}=-\partial_{[\mu}\beta_{\nu]}=\text{const}.
  \end{align}
  As is shown clearly in Eq.(\ref{condition}) , the spin potential $\Omega_{\mu\nu}$  is fixed to thermal vorticity $\frac{1}{2}(\partial_\nu \beta_\mu-\partial_\mu \beta_\nu)$, and $\beta^\mu$ can be further  decomposed into a translation ($a^\mu$) and a rigid rotation ($\Omega^{\mu\nu}x_\nu$)  in global equilibrium,
  \begin{align}
  &\beta^\mu=a^\mu+\Omega^{\mu\nu}x_\nu,\quad a^\mu=\text{const},
  \end{align}
  which are consistent with the previous conclusions drawn in  \cite{Becattini:2013fla,Florkowski:2017ruc}.

   \section{Motion equations of hydrodynamic variables}
   \label{otion}	
  In this section, we derive the motion equations of all relevant hydrodynamic variables.
  As is know to all of us, hydrodynamics is based on macroscopic conservation laws,
  \begin{align}
  \label{ncon}
  \partial_\mu N^\mu&=0,\\
  \label{Tcon}
  \partial_\mu T^{\mu\nu}&=0,\\
  \label{scon}
  \partial_\lambda S^{\lambda,\mu\nu}&=2T^{[\nu\mu]},
  \end{align}
  where aside from charge and energy-momentum we consider also total angular momentum conservation when it comes to spin hydro. We remind the readers that the conserved current $N^\mu$ is trivial particle number current for we take only two body scatterings into account in the sector of kinetics. 
  
  To put the above equations less abstact, we identify the definitions of the parameters $\mu(x),T(x),u^\mu(x)$, i.e, $n(x),e(x),u^\mu(x)$ first. We assume that the particle number density and  energy density are completely determined by the local equilibrium distribution function $f^{(0)}$, 
  which, with the identification of fluid velocity, amounts to the Landau matching conditions:
  \begin{align}
  &n\equiv\int d\Gamma \,p^\mu u_\mu f(x,p,\bm{s})=\int d\Gamma \,p^\mu u_\mu f^{(0)}(x,p,\bm{s}), \\
  &e\equiv\int d\Gamma\,(p^\mu u_\mu)^2 f(x,p,\bm{s})=\int d\Gamma \,(p^\mu u_\mu)^2 f^{(0)}(x,p,\bm{s}),
  \end{align}
  and we adopt Landau velocity $T^{\mu\nu}u_\nu=eu^\mu$.   Here $f^{(0)}$ represents the one-particle distribution in local equilibrium and this expansion is based on small spin potential $\Omega$,
  \begin{align}
  & f^{(0)}=(1+\frac{\Omega_{\mu\nu}\Sigma_{\bm{s}}^{\mu\nu}}{4})f^{(0)}_{\text{ws}},\\
  & f^{(0)}_{\text{ws}}=\frac{1}{(2\pi)^3}\exp[\xi-\beta\cdot p],
  \end{align}
  where $f^{(0)}_{\text{ws}}$ denotes the spinless distribution function in local equilibrium.
  
  The densities $n$ and $e$ together with the static pressure $P$
  \begin{align}
  &P\equiv-\frac{1}{3}\int d\Gamma \,\Delta_{\mu\nu}p^\mu p^\mu f^{(0)}(x,p,\bm{s}),
  \end{align}
  are analytically evaluated using the formulas in Appendix.(\ref{int}),
  \begin{align}
  &n=\exp(\xi)n_0(T)=\exp(\xi)\frac{T^3}{2\pi^2}z^2 K_2(z),\nn\\
  &e=\exp(\xi)e_0(T)=\exp(\xi)\frac{T^4}{2\pi^2}z^2 (3K_2(z)+zK_1(z)),\nn\\
  &P=\exp(\xi)n_0(T)T,
  \end{align}
  with $z\equiv\frac{m}{T}$.

  Noting that in local equilibrium, we have:
  \begin{align}
  \label{N0}
  &N^{(0)\mu}=\int\,d\Gamma \,p^\mu f^{(0)}=nu^\mu,\\
  \label{T0}
  &T^{(0)\mu\nu}=\int \,d\Gamma \,p^\mu p^\nu f^{(0)}=eu^\mu u^\nu-P\Delta^{\mu\nu},\\
  \label{S0}
&S^{(0)\lambda,\mu\nu}=\int \,d\Gamma \,p^\lambda (\frac{1}{2}\Sigma^{\mu\nu}_{\bm{s}}-\frac{1}{2m^2}p^{[\mu}\partial^{\nu]} )f^{(0)}\nn\\
&\quad\quad\quad\,\;=\frac{\exp(\xi)}{4}(n_0(T)u^\lambda\Omega^{\mu\nu}+(\frac{6}{z^2}\frac{e_0(T)+P_0(T)}{T}+2n_0(T)\,)u^\lambda u^\delta  u^{[\mu}\Omega^{\nu]}_{\,\,\delta}\nn\\
&\quad\quad\quad\,\;-\frac{2}{z^2}\frac{e_0(T)+P_0(T)}{T}(\Delta^{\lambda\delta}u^{[\mu}  \Omega^{\nu]}_{~\delta}
+ u^\lambda \Delta^{\delta[\mu} \Omega^{\nu]}_{~\delta}
+ u^\delta \Delta^{\lambda[\mu} \Omega^{\nu]}_{~\delta})\,)\nn\\
&\quad\quad\quad\,\;-\frac{\exp(\xi)}{2m^2}\big(eu^\lambda u^{[\mu}\partial^{\nu]}\xi-(I_{30}u^\lambda u^\rho u^{[\mu}+I_{31}u^\lambda\Delta^{\rho[\mu})\,\partial^{\nu]}\beta_\rho\big)\nn\\
&\quad\quad\quad\,\;+\frac{\exp(\xi)}{2m^2}\big(\,P\Delta^{\lambda[\mu}\partial^{\nu]}\xi+I_{31}(\Delta^{\lambda\rho}u^{[\mu}+\Delta^{\lambda[\mu}u^\rho)\partial^{\nu]}\beta_\rho\,\big).
  \end{align}
  
  Combining Eqs.(\ref{ncon}), (\ref{Tcon}), (\ref{N0}) and (\ref{T0}) with the enthalpy $h\equiv e+P$, we find that:
  \begin{align}
  \label{ME1}
  &u_\nu\partial_\mu T^{\mu\nu}=(e+P)\theta+u\cdot\partial e=0,  \\
  \label{ME2}
  &\Delta_{\alpha\nu}\partial_\mu T^{\mu\nu}=(e+P)Du_\alpha-\partial_\alpha P+ u_\alpha(u\cdot\partial P)=0,
  \end{align}
  and 
  \begin{align}
  \label{Dn}
  &Dn=-n\theta,  \\
  \label{De}
  &De=-(e+P)\theta,  \\
  \label{Du}
  &Du_\alpha=\frac{1}{h}\nabla_\alpha P .
  \end{align}
  And Eqs.(\ref{Dn}) and (\ref{De}) can be further transformed into the lowest order evolution equations for $\beta$ and $\xi$  
  \begin{align}
  \label{DT}
  &D\beta=\frac{1}{D_{20}}(-I_{20}n\theta+I_{10}h\theta), \\
  \label{Dalpha}
  &D\xi=\frac{1}{D_{20}}(-I_{30}n\theta+I_{20}h\theta),
  \end{align}
  where the definitions of  thermodynamic integrals $I_{nq}$ and $D_{nq}$ are put in Appendix.\ref{int}. One of fascinating features of spin hydrodynamic theory is that  it can describe the relaxation of spins, therefore it is important to make out the motion equation of spin potential, namely, $D\Omega^{\mu\nu}$. On the other hand, $D\Omega^{\mu\nu}$ has to be substituted by known functions and their gradients in order to seek solutiuons to  the  linearized Boltzmann equation in the next section. The
  motion equation for $\Omega$  has been formulated by some theoretical works. Similar to the results in \cite{Peng:2021ago} and \cite{Bhadury:2020cop}, we now formulate this equation of motion on the basis of collisionless case.  It is clear that our definitions for both energy-momentum tensor and spin tensor receive no corrections from collisional effects now. As a supplement, we note that lack of nonlocality in the collision operator of relaxation time approximation (RTA) \cite{Bhadury:2020cop} leads to conservation of spin angular momentum and symmetry of energy-momentum tensor.
  In fact, collisions do contribute to $D\Omega^{\mu\nu}$ via the anti-symmetric part of $T^{\mu\nu}$.
   Because nonlocal collisions are explicitly included in Eq.(\ref{cf}), the anti-symmetric part of energy-momentum tensor arises in the order of $O(\partial^2)$ as a result of collisional effects \cite{Weickgenannt:2020aaf}, but this is irrelevant to present first-order construction,   which also means that the mechanism of spin-orbit coupling does not come into play in first-order spin hydrodynamics. In this section, we derive the motion equation for $\Omega$ via only 
  \begin{align}
  \label{sconserv}
  \partial_\lambda S^{(0)\lambda,\mu\nu}=2T^{(0)[\nu\mu]}=0.
  \end{align}

   
  Then substitution of Eq.(\ref{S0}) into the above equation leads to 
  \begin{align}
  \label{Do}
  D\Omega ^{\mu \nu}=& R_{\pi }^{[\nu }{}_{\lambda }\sigma ^{\lambda\mu ]}+R_{\Pi }^{\mu \nu }\theta +\left(\nabla ^{\alpha }\xi \right)R_n^{[\mu \nu ]}{}_{\alpha }+R_{\text{$\Sigma $1}}^{\alpha }\nabla ^{[\mu }\Omega ^{\nu ]}{}_{\alpha }+R_{\text{$\Sigma $2}}^{[\mu \nu ]\alpha }\nabla ^{\lambda }\Omega _{\alpha \lambda },
  \end{align}
  where the coefficients are put in Appendix.(\ref{do}) and similar results can be also found in \cite{Bhadury:2020cop} with only differences in the factor of $\xi$.  Here we discard the terms of $\text{O}(\partial^2)$ and keep only $\text{O}(\partial \Omega)$ and $\text{O}(\Omega\partial)$.  Although $\Omega$ is deemed as $O(\partial)$ as far as its magnititude of order is concerned, its status is enhanced to a basic hydrodynamic variable thus we retain the gradients of $\Omega$ and $O(\Omega\partial)$  in first-order theory. 
  \clearpage
 
 \section{Chapman-Enskog expansion}
  \label{expansion}	
Before moving on, we take some time to introduce our expansion scheme. Although we title this section with "Chapman-Enskog expansion", there are actually two expansion countings in presence, one is authorized Chapman-Enskog expansion featured with mean free path $l_{\text{mfp}}$ that is also called Knudsen number expansion with the expansion parameter $K_n\equiv\frac{l_\text{mfp}}{L_{\text{hydro}}}$ ($L_{\text{hydro}}$ is characteristic length associated with system non-uniformity), the other one is spin expansion characterized by new scale $\Delta$ (non-local shift). In order for the assumption of molecular chaos to hold, these two scales should satisfy \cite{Weickgenannt:2021cuo}
\begin{align}
\Delta \lesssim l_{\text{mfp}}.
\end{align}
In addition, small spin potential expansion is also adopted. If various expansion schemes all bring in new characteristic length scales, it may cause confusion when counting order.
 For simplicity, we require that these mensioned length scales are close to each other so that the distinctions  needn't  be made, which greatly facilitates our investigation.

 To proceed, we follow the spirit of Chapman-Enskog expansion and  one can find the related details from any textbooks on kinetic theory. First, we have,
\begin{align}
\label{boltz1}
&p\cdot u Df+\epsilon p^\nu\nabla_\nu f=C[f],
\end{align}
and the follow expansion is employed
\begin{align}
\label{fk}
&f=f^{(0)}+\epsilon f^{(1)}+...   \\
\label{Dfk}
&Df=\epsilon(Df)^{(1)}+...  \\
&f^{(1)}=f^{(0)}\phi ,
\end{align}
 where the  function $\phi$ characterizes the deviation of realistic distribution away from reference local equilibrium distribution. 
  A book-keeping parameter $\epsilon$ is introduced measuring the relative strength of the gradients, which is called the non-uniformity parameter and is  identified as the well-known Knudsen number. 
 
   To solve the linearized Boltzmann equation, we also need to refer to the solubility conditions, which stem from the conservation laws of particle number and energy-momentum and read that $1$ and $p^\mu$ are collosional invariants,
   \begin{align}
   \label{nconserv}
   &\int d\Gamma \,p^\mu u_\mu(D f)^{(1)}=-\int d\Gamma \,p^\mu \nabla_\mu f^{(0)}, \\
   \label{Tconserv}
   &\int d\Gamma\,p^\mu p^\nu u_\nu(D f)^{(1)}=-\int d\Gamma\,p^\mu p^\nu \nabla_\nu f^{(0)}.
   \end{align} 
   The above two equations are exactly Eqs.(\ref{ncon}) and (\ref{Tcon}) of kinetic version regardless of the superscripts labeling order and will give Eqs.(\ref{DT}) and (\ref{Dalpha}), too. Physically, there should be another collisional invariant $J^{\mu\nu}$ tightly connected with Eq.(\ref{scon}), and the conditions in Eq.(\ref{condition}) for global equilibrium rely on the assumption of conserved total angular momentum (the collsion kernel itself does not conserve $J^{\mu\nu}$ for lack of a delta function like $\delta^{(4)}(p+p^\prime-p_1-p_2)$). Recent work about the analysis on sound propogation and spin relaxation has phenomenologically incorporated this point \cite{Hu:2022lpi}, but we here hold the collision term unchanged and choose to calculate the motion equation for $\Omega$ in hydro sector done in the previous section.
   
   To proceed, we write $(D f)^{(1)}$ as:
   \begin{align}
   (D f)^{(1)}=\frac{\partial f^{(0)}}{\partial \xi}(D \xi)^{(1)}+\frac{\partial f^{(0)}}{\partial \beta}(D \beta)^{(1)}+\frac{\partial f^{(0)}}{\partial u^\mu}(D u^\mu)^{(1)}+\frac{\partial f^{(0)}}{\partial \Omega^{\mu\nu}}(D \Omega^{\mu\nu})^{(1)}.
   \end{align}

By power counting and setting $\epsilon$ one, Eq.(\ref{boltz1}) reduces to,
\begin{align}
\label{boltz2}
&p\cdot u (Df)^{(1)}+ p^\nu\nabla_\nu f^{(0)}=L[\phi]+L[\chi_1]+L[\chi_2]+L[\chi_3]+L[\chi_4]+L[\chi_5],
\end{align}
with
\begin{align}
L[\phi]
&\equiv \frac{1}{(2\pi)^6} \int d\Gamma^\prime d\Gamma_1 d\Gamma_2  \,
\mathcal{W} \,  \exp(2\xi-\beta\cdot (p+p^\prime)\,)  \nn\\
&\times\big( \phi(x,p_1,\bm{s}_1) +\phi(x,p_2,\bm{s}_2) -\phi(x,p,\bm{s})-\phi(x,p^\prime,\bm{s}^\prime) \big),\\
\chi_1&\equiv-\partial_\mu\beta_\nu \Delta^\mu p^\nu+\partial_\mu\xi \Delta^\mu,\quad \chi_2\equiv\frac{1}{4}\Omega_{\mu\nu}\Sigma_{\bm{s}}^{\mu\nu},\nn\\
\chi_3&\equiv\frac{1}{4}\partial_\rho\Omega_{\mu\nu}\Delta^\rho\Sigma_{\bm{s}}^{\mu\nu},\quad \chi_4\equiv\frac{1}{4}\Omega_{\mu\nu}\Sigma_{\bm{s}}^{\mu\nu}\phi^{(1)}(x,p,\bm{s}),\\
\chi_5 &=\frac{1}{4}\Omega_{\rho\sigma}\Sigma_{\bm{s}}^{\rho\sigma}(-\partial_\mu\beta_\nu \Delta^\mu p^\nu+\partial_\mu\xi \Delta^\mu),
\end{align}
and
\begin{eqnarray}
(Df)^{(1)}&=&(1+\frac{\Omega_{\mu\nu}\Sigma_{\bm{s}}^{\mu\nu}}{4})f^{(0)}_{\text{ws}}(D\xi)^{(1)}-p\cdot u(1+\frac{\Omega_{\mu\nu}\Sigma_{\bm{s}}^{\mu\nu}}{4})f^{(0)}_{\text{ws}}(D\beta)^{(1)}-\beta(1+\frac{\Omega_{\mu\nu}\Sigma_{\bm{s}}^{\mu\nu}}{4})f^{(0)}_{\text{ws}}p_\alpha(Du^\alpha)^{(1)}\nn\\
&+&\frac{\Sigma_{\bm{s}}^{\alpha\beta}}{4}f^{(0)}_{\text{ws}}(D\Omega_{\alpha\beta})^{(1)},\nn\\
\nabla_\alpha f^{(0)}&=&(1+\frac{\Omega_{\mu\nu}\Sigma_{\bm{s}}^{\mu\nu}}{4})f^{(0)}_{\text{ws}}\nabla_\alpha\xi-p\cdot u(1+\frac{\Omega_{\mu\nu}\Sigma_{\bm{s}}^{\mu\nu}}{4})f^{(0)}_{\text{ws}}\nabla_\alpha\beta-\beta(1+\frac{\Omega_{\mu\nu}\Sigma_{\bm{s}}^{\mu\nu}}{4})f^{(0)}_{\text{ws}}p_\beta\nabla_\alpha u^\beta\nn\\
&+&\frac{\Sigma_{\bm{s}}^{\beta\gamma}}{4}f^{(0)}_{\text{ws}}\nabla_\alpha\Omega_{\beta\gamma},
\end{eqnarray}
where $\phi=\phi^{(1)}+\phi^{(2)}$ according to order seperation and we have neglected cross terms in both $L[\chi_4]$ and $L[\chi_5]$, for example,  the terms $\Sigma_{\bm{s}_1}^{\mu\nu}\phi^{(1)}(x,p_2,\bm{s}_2)$ and $\Sigma_{\bm{s}_2}^{\mu\nu}\phi^{(1)}(x,p_1,\bm{s}_1)$ are all neglected. This approximation can be understood by arguing that the derivative expansion $\Delta\cdot\partial$  and  spin potential  expansion are both implemented on the same distribution function and this principle is also applied to choose the combination of $\Omega_{\mu\nu}\Sigma_{\bm{s}}^{\mu\nu}$ and $\phi^{(1)}$ in $L[\chi_4]$. With this constriant, two collision terms on discussion are all cast into a uniform form like $L[\phi]$ just as shown at present, which will greatly simplify our follow-up calculations. (one can surely include discarded cross terms for completeness, but that will introduce more new tensor structures). 
 In the order of $\text{O}(1)$, $f^{(0)}_{\text{ws}}$ must be substituted into $C[f]$ and $C[f]$ vanishes. 

When moving to 
 higher order of $\text{O}(\partial)$ and $\text{O}(\partial\Omega)$, we obtain,
\begin{align}
\label{part}
&(D\xi)^{(1)}-p\cdot u(D\beta)^{(1)}-\beta p_\alpha(Du^\alpha)^{(1)}+\nabla_\alpha\xi-p\cdot u\nabla_\alpha\beta-\beta p_\beta\nabla_\alpha u^\beta=\frac{L[\chi_1]+L[\chi_2]+L[\phi^{(1)}]}{f^{(0)}_{\text{ws}}},\\
\label{o}
&\Omega_{\mu\nu}\Sigma_{\bm{s}}^{\mu\nu}\big[(D\xi)^{(1)}-p\cdot u(D\beta)^{(1)}-\beta p_\alpha(Du^\alpha)^{(1)}+\nabla_\alpha\xi-p\cdot u\nabla_\alpha\beta-\beta p_\beta\nabla_\alpha u^\beta\big]\nn\\
&\quad\quad+p^\alpha\Sigma_{\bm{s}}^{\beta\gamma}\nabla_\alpha\Omega_{\beta\gamma}+\Sigma_{\bm{s}}^{\alpha\beta}(D\Omega_{\alpha\beta})^{(1)}=\frac{4(L[\chi_3]+L[\chi_4]+L[\chi_5]+L[\phi^{(2)}])}{f^{(0)}_{\text{ws}}},
\end{align}
where the deviation function $\phi$  balancing above two equations is left undetermined and $(DF)^{(1)}$ in kinetic sector corresponds to the motion equation $DF$ in hydrodynamic sector ($F$ repesents $\xi$, $\beta$, $u$ or $\Omega$).
For solving Eq.(\ref{part}), the first-order deviation function $\phi$ can be conveniently chosen as 
\begin{align}
\label{phi1o}
\phi^{(1)}(x,p,\bm{s})&=\phi_1^{(1)}(x,p,\bm{s})+\phi_0^{(1)}(x,p),\\
\label{phi11}
\phi_1^{(1)}(x,p,\bm{s})&=\partial_\mu\beta_\nu \Delta^\mu p^\nu-\partial_\mu\xi \Delta^\mu-\frac{1}{4}\Omega_{\mu\nu}\Sigma_{\bm{s}}^{\mu\nu}, \\
\label{phi12}
\phi_0^{(1)}(x,p)&=\frac{1}{n\sigma(T)}(A(\tau,z)\theta+\beta B(\tau,z)p^{\langle\alpha\rangle}\nabla_\alpha\xi+\beta^2C(\tau,z)p^{\langle\alpha}p^{\beta\rangle}\sigma_{\alpha\beta }),
\end{align}
where we have invoked that $\phi_1^{(1)}(x,p,\bm{s})$ cancels the other two collision kernels in Eq.(\ref{part}),  $\phi_0^{(1)}(x,p)$  characterizes the solution to spinless linearized Boltzmann equation with the same notaions as used in \cite{DeGroot:1980dk}, which follows from
\begin{align}
\label{part1}
&(D\xi)^{(1)}-p\cdot u(D\beta)^{(1)}-\beta p_\alpha(Du^\alpha)^{(1)}+\nabla_\alpha\xi-p\cdot u\nabla_\alpha\beta-\beta p_\beta\nabla_\alpha u^\beta=\frac{L[\phi_0^{(1)}]}{f^{(0)}_{\text{ws}}},
\end{align}
 $\sigma(T)$ is an arbitrary constant with the dimension of cross sections, and $\tau\equiv\beta u\cdot p$. Noticing that Eq.(\ref{part1}) has been extensively solved for these dimensionless functions $A(\tau,z),B(\tau,z)$, and $C(\tau,z)$ via different 
methods with various interactions, we here see $\phi_0^{(1)}$ as a known function.

One may keenly observe that $L[\phi_0^{(1)}]$ is distinct from the linearized collision kernel $\mathcal{L}[\phi]$ exhibited in \cite{DeGroot:1980dk} on account of various transition rates. This can be understood by rethinking that spin must be integrated out (averaged)  for lack of spin information in left hand side (LHS) of Eq.(\ref{part1}), which indicates that we are dealing with spinless hydro after averaging spin in both sides. Therefore, $\int dS(p)L[\phi_0^{(1)}]$ recovers familiar form of $\mathcal{L}[\phi]$. Together with the motion equations we get in the preceding section, we conclude that the sector of spinless hydrodynamics is successfully recovered and dose not vary for including spin, which is reasonable considering its universality as
a low-energy effective theory.

By substituting Eq.(\ref{phi1o}) into $L[\chi_4]$ and one find that part of $L[\chi_4]$ cancels $L[\chi_5]$ (the terms of $O(\Omega^2)$ are omitted), 
\begin{align}
\label{o0}
&\Omega_{\mu\nu}\Sigma_{\bm{s}}^{\mu\nu}\big[(D\xi)^{(1)}-p\cdot u(D\beta)^{(1)}-\beta p_\alpha(Du^\alpha)^{(1)}+\nabla_\alpha\xi-p\cdot u\nabla_\alpha\beta-\beta p_\beta\nabla_\alpha u^\beta\big]\nn\\
&\quad\quad+p^\alpha\Sigma_{\bm{s}}^{\beta\gamma}\nabla_\alpha\Omega_{\beta\gamma}+\Sigma_{\bm{s}}^{\alpha\beta}(D\Omega_{\alpha\beta})^{(1)}=4f^{-1(0)}_{\text{ws}}(L[\chi_3]+L[\frac{1}{4}\Omega_{\mu\nu}\Sigma_{\bm{s}}^{\mu\nu}\phi_0^{(1)}(x,p)]+L[\phi^{(2)}]).
\end{align}

Similarly, $\phi^{(2)}$ can be parameterized as
\begin{align}
\label{phi20}
\phi^{(2)}(x,p,\bm{s})&=\phi_1^{(2)}(x,p,\bm{s})+\phi_0^{(2)}(x,p,\bm{s}),\\
\label{phi21}
\phi_1^{(2)}(x,p,\bm{s})&=-\frac{1}{4}\partial_\rho\Omega_{\mu\nu}\Delta^\rho\Sigma_{\bm{s}}^{\mu\nu}-\frac{1}{4}\Omega_{\mu\nu}\Sigma_{\bm{s}}^{\mu\nu}\phi_0^{(1)}(x,p),
\end{align}
to counteract the other two collision terms in Eq.(\ref{phi20}).
After cancellation of three linearized collision terms, Eq.(\ref{o0}) turns into
\begin{align}
\label{o1}
&\Omega_{\mu\nu}\Sigma_{\bm{s}}^{\mu\nu}\big[(D\xi)^{(1)}-p\cdot u(D\beta)^{(1)}-\beta p_\alpha(Du^\alpha)^{(1)}+\nabla_\alpha\xi-p\cdot u\nabla_\alpha\beta-\beta p_\beta\nabla_\alpha u^\beta\big]\nn\\
&\quad\quad+p^\alpha\Sigma_{\bm{s}}^{\beta\gamma}\nabla_\alpha\Omega_{\beta\gamma}+\Sigma_{\bm{s}}^{\alpha\beta}(D\Omega_{\alpha\beta})^{(1)}=\frac{4L[\phi_0^{(2)}]}{f^{(0)}_{\text{ws}}}.
\end{align}
 Considering the resemblance in Eqs.(\ref{part1}) and (\ref{o1}), solving Eq.(\ref{o1}) can be done in a similar manner to that employed to solve Eq.(\ref{part1}) in \cite{DeGroot:1980dk}.
To solve Eq.(\ref{o1}),  we substitute all temporal derivatives obtained in the previous section and expand $\nabla_\alpha\Omega_{\beta\gamma}$ like
\begin{align}
\nabla_\alpha\Omega_{\beta\gamma}=\nabla_{[\alpha}\Omega_{\beta]\gamma}+\nabla_{\langle\alpha}\Omega_{\beta\rangle\gamma}+\frac{1}{3}\Delta_{\alpha\beta}\nabla^\lambda\Omega_{\lambda\gamma}+u^\lambda u_{(\alpha}\nabla_{\beta)}\Omega_{\lambda\gamma},
\end{align}
then the second order deviation function $\phi^{(2)}_0$ can be parameterized according to the LHS of Eq.(\ref{o1}),
\begin{align}
\label{phi2}
\phi_0^{(2)}&=\frac{1}{n\sigma(T)}\big[\big(A_1(\tau,z)\Sigma_{\bm{s},\rho\sigma}\Omega^{\rho\sigma}+A_2(\tau,z)\Sigma_{\bm{s},\rho\sigma}u^{\alpha} u^{[\rho} \Omega^{\sigma]}{}_{\alpha}\big)\theta\nn\\
&+\big(\beta B_1(\tau,z)\Sigma_{\bm{s},\rho\sigma}\Omega^{\rho\sigma}p_{\alpha}+B_2(\tau,z)\Sigma_{\bm{s},\rho\sigma}u^{[\rho}\Omega^{\sigma]}_{\;\;\alpha}+B_3(\tau,z)\Sigma_{\bm{s},\rho\sigma}u^{\kappa}g^{[\rho}{}_{\alpha}  \Omega^{\sigma]}{}_{\kappa}\big)\nabla^\alpha\xi\nn\\
&+\big(\beta^2C_1(\tau,z)\Sigma_{\bm{s},\rho\sigma}\Omega^{\rho\sigma }p_{\langle\alpha}p_{\beta\rangle}+C_2(\tau,z)\Sigma_{\bm{s},\rho\sigma}\Omega^{[\rho}_{\;\;\beta} g^{\sigma]}_{\;\;\alpha}+C_3(\tau,z)\Sigma_{\bm{s},\rho\sigma}u^{[\rho} g^{\sigma]}_{\;\;\alpha}u^\kappa \Omega_{\kappa\beta}\big)\sigma^{\alpha\beta }\nn\\
&+( F_1(\tau,z)\Sigma_{\bm{s}}^{\mu\nu}u_{\alpha}+\beta F_2(\tau,z)p^\mu\Sigma_{\bm{s},\alpha}^\nu )\nabla_{[\mu} \Omega_{\nu]}^{\,\;\;\alpha}+\big( G_1(\tau,z)\Sigma_{\bm{s},\rho\sigma}u^{[\rho} g^{\sigma ]\gamma }+\beta G_2(\tau,z)\Delta_{\alpha\beta}p^{\alpha}\Sigma_{\bm{s}}^{\beta\gamma}\,\big)\nabla^{\lambda} \Omega_{\lambda\gamma}\nn\\
&+\beta H(\tau,z)p^\mu\Sigma_{\bm{s}}^{\rho\sigma}\nabla_{\langle\mu} \Omega_{\rho\rangle\sigma}+\beta J(\tau,z)p^\alpha\Sigma_{\bm{s}}^{\rho\sigma}u^\lambda u_{(\alpha}\nabla_{\rho)}\Omega_{\lambda\sigma}\big],
\end{align}
where  $A_i,B_i,C_i,F_i,G_i, H,J$ ($i=1, 2$ or 3) are dimensionless functions  to be determined by solving corresponding integral equations.

 Considering all the thermodynamic forces relevant including the gradients of spin potential we explicitly write above are independent, this highly involved equation can be solved by equating the coefficients of each of the thermodynamic forces separately,  
  thus the solution to Eq.(\ref{o1}) is equivalent to solving the set of equations. However, even corresponding to one unique thermodynamic force, different terms are also independent, which means that the procedure of seperate equation can be enlarged to every term shown in Eq.(\ref{phi2}). For instance, the integral equations for the terms related to  $\sigma^{\alpha\beta}$ are expressed as
\begin{align}
&\beta^2\Sigma_{\bm{s}}^{\rho\sigma}\Omega_{\rho\sigma}p^{\langle\alpha}p^{\beta\rangle}=\frac{4C_\phi}{f_{\text{ws}}}\{\phi=\frac{\beta^2}{n\sigma(T)}\Sigma_{\bm{s}}^{\rho\sigma}\Omega_{\rho\sigma}p^{\langle\alpha}p^{\beta\rangle}C_1(\tau,z)\} \quad,\\
&\frac{4 I_{31}}{(m^2\, I_{10} - 2 I_{31})} \Sigma_{\bm{s},\rho\sigma}\Omega^{[\rho}_{\;\;\beta} g^{\sigma]}_{\;\;\alpha}=\frac{4C_\phi}{f_{\text{ws}}}\{\phi=\frac{1}{n\sigma(T)}\Sigma_{\bm{s},\rho\sigma}\Omega^{[\rho}_{\;\;\beta} g^{\sigma]}_{\;\;\alpha}C_2(\tau,z)\} \quad,\\
&\frac{4 (I_{30} - I_{31}) I_{31}\Sigma_{\bm{s},\rho\sigma}u^{[\rho} g^{\sigma]}_{\;\;\alpha}u^\kappa \Omega_{\kappa\beta}}{(m^2\, I_{10} - 2\, I_{31}) \big[m^2\, I_{10} - (I_{30} + I_{31})\big]}=\frac{4C_\phi}{f_{\text{ws}}}\{\phi=\frac{1}{n\sigma(T)}\Sigma_{\bm{s},\rho\sigma}u^{[\rho} g^{\sigma]}_{\;\;\alpha}u^\kappa \Omega_{\kappa\beta}C_3(\tau,z)\} \quad,
\end{align}
where we have prescibed that the notation $F[\phi]\{\phi=A\}$ represent that  $\phi$ is taken to be $A$ in functional $F$. Another equations follow from repetitive practice and are not shown here. 

Before ending this section, there are some comments. In the process of seeking a solution to Eq.(\ref{part1}), we point out that  only by averaging spin can normal hydrodynamic equations be recovered for spin dependence mismatching in both sides. Things are the same when  handing spin sector that the LHS of Eq.(\ref{o1}) has only one $\bm{s}$ while the RHS has two spin sources $\phi(x,p,\bm{s})$ and $\mathcal{W}$. Following the spirit of recent related review work \cite{Hidaka:2022dmn}, the equality relation Eq.(\ref{o1}) should be loosely understood as equivalence after spin integration $\int dS(p)\bm{s}$, which equivalently suggests that spin-related observables are defined in this way with classical interpretation for spin. Last but not the least, those equations to be solved reduces to momentum dependent only after spin integration, therefore, well-developped methods for solving or approximately solving linearized transport equation can be sufficiently employed.  

 \section{Nonequilibrium corrections to $N^\mu$, $T^{\mu\nu}$, and $S^{\lambda,\mu\nu}$}
 \label{corr}	
 In the previous section, the formal solution to deviation function $\phi$ is determined, which is composed of four parts given in Eqs.(\ref{phi11}), (\ref{phi12}), (\ref{phi21}) and (\ref{phi2}). From its expression, the corrections from nonlocal effects are included, which are marked by $\Delta^{\mu}$. As a reminder, the distribution function $f(x,p,\bm{s})$ appears inside the definitions of tensors such as Eqs.(\ref{N}), (\ref{T}) and (\ref{S}) instead of $f(x+\Delta,p,\bm{s})$. Thus when it comes to nonequilibrium corrections to $N^\mu$, $T^{\mu\nu}$, and $S^{\lambda,\mu\nu}$, the $\Delta$ dependence has to be removed by hand. So in this section $\phi=\phi|_{\Delta^\mu=0}$.
 
Here and now the corrections to $T^{\mu\nu}$ and $N^{\mu}$ can be persued with the formal solution of $\phi$ presented in Sec.\ref{expansion}, 
\begin{align}
\label{N1}
&N^{\mu}=\int\,d\Gamma \,p^\mu f^{(0)}(1+\phi)=nu^\mu+V^\mu,\\
\label{T1}
&T^{\mu\nu}=\int \,d\Gamma \,p^\mu p^\nu f^{(0)}(1+\phi)=eu^\mu u^\nu-P\Delta^{\mu\nu}+\pi^{\mu\nu}+\Pi\Delta^{\mu\nu},
\end{align}
where the dissipative quantities $V^\mu, \pi^{\mu\nu}$ and $\Pi$ can all be obtained by projection to $N^\mu$ and $T^{\mu\nu}$. One can observe that these expressions show no difference from the ordinary first-order viscous fluids owing to Eq.(\ref{f2}), which means no corrections up to $\text{O}(\partial\Omega)$ and $\text{O}(\Omega\partial)$. 

 It is well-known that first-order viscous hydrodynamics is characterized by linear laws between dissipative quantities and thermodynamic forces. These can be achieved following the practice in textbooks, we write with no detailed derivation
\begin{align}
\pi^{\mu\nu}=2\eta\sigma^{\mu\nu},\\
\Pi=\zeta\theta,\\
V^\mu=\kappa\nabla^\mu\xi,
\end{align}
where $\eta,\zeta,\kappa$ correspond to shear viscosity, bulk viscosity and diffusion coefficient respectively.
Solving the integral functions for $A_1,B_1,C_1$ and obtain all the transport coefficients can be formulated in a systematic way \cite{DeGroot:1980dk}. 

In the end of this section, we evaluate the corrections to spin tensor $\delta S^{\lambda,\mu\nu}$. The full spin tensor can be split into
\begin{align}
\label{S1}
&S^{\lambda,\mu\nu}=S^{(0)\lambda,\mu\nu}+\delta S^{\lambda,\mu\nu},
\end{align}
and the second part originates from the nonequilibrium deviation $\phi$
\begin{align}
\label{delta2}
&\quad\quad\quad\,\;\delta S^{\lambda,\mu\nu}=\frac{1}{2}\int \,d\Gamma \,p^\lambda \Sigma^{\mu\nu}_{\bm{s}}\exp[\xi-\beta\cdot p]\phi^{(2)}(x,p,\bm{s})+\frac{1}{8}\int \,d\Gamma \,p^\lambda \Sigma^{\mu\nu}_{\bm{s}}\Sigma_{\bm{s}}^{\rho\sigma}\Omega_{\rho\sigma}\exp[\xi-\beta\cdot p]\phi^{(1)}(x,p,\bm{s})\nn\\
&\quad\quad\quad\quad\quad\,\;\quad=\frac{1}{2}\int \,d\Gamma \,p^\lambda \Sigma^{\mu\nu}_{\bm{s}}\exp[\xi-\beta\cdot p]\big(\phi_0^{(2)}(x,p,\bm{s})-\frac{1}{4}\Sigma_{\bm{s}}^{\rho\sigma}\Omega_{\rho\sigma}\phi_0^{(1)}(x,p)\big)\nn\\
&\quad\quad\quad\quad\quad\,\;\quad+\frac{1}{8}\int \,d\Gamma \,p^\lambda \Sigma^{\mu\nu}_{\bm{s}}\Sigma_{\bm{s}}^{\rho\sigma}\Omega_{\rho\sigma}\exp[\xi-\beta\cdot p]\phi_0^{(1)}(x,p,\bm{s}),
\end{align}
where Eq.(\ref{f31}) has been utilized in the second equality and the second term in Eq.(\ref{S}) contributes only $O(\partial^2)$ thus is discarded. To make the expression consise, an auxiliary tensor $X^{\rho\sigma}$ is introduced,
\begin{align}
X^{\rho\sigma}&=\frac{1}{n\sigma(T)}\big[\big(A_1(\tau,z)\Omega^{\rho\sigma}+A_2(\tau,z)u^{\alpha} u^{[\rho} \Omega^{\sigma]}{}_{\alpha}\big)\theta\nn\\
&+\big(\beta B_1(\tau,z)\Omega^{\rho\sigma}p_{\alpha}+B_2(\tau,z)u^{[\rho}\Omega^{\sigma]}_{\;\;\alpha}+B_3(\tau,z)u^{\kappa}g^{[\rho}{}_{\alpha}  \Omega^{\sigma]}{}_{\kappa}\big)\nabla^\alpha\xi\nn\\
&+\big(\beta^2C_1(\tau,z)\Omega^{\rho\sigma }p_{\langle\alpha}p_{\beta\rangle}+C_2(\tau,z)\Omega^{[\rho}_{\;\;\beta} g^{\sigma]}_{\;\;\alpha}+C_3(\tau,z)u^{[\rho} g^{\sigma]}_{\;\;\alpha}u^\kappa \Omega_{\kappa\beta}\big)\sigma^{\alpha\beta }\nn\\
&+F_1(\tau,z)u_{\alpha} \nabla^{[\rho} \Omega^{\sigma]\alpha}+\beta F_2(\tau,z)p_\mu\nabla^{[\mu} \Omega^{\rho]\sigma}+ G_1(\tau,z)u^{[\rho} g^{\sigma ]\gamma }\nabla^{\lambda} \Omega_{\lambda\gamma}+\beta G_2(\tau,z)\Delta^{\alpha\rho}p_{\alpha}\nabla_{\lambda} \Omega^{\lambda\sigma}\nn\\
&+\beta H(\tau,z)p_\mu\nabla^{\langle\mu} \Omega^{\rho\rangle\sigma}+\beta J(\tau,z)p_\alpha u_\lambda u^{(\alpha}\nabla^{\rho)}\Omega^{\lambda\sigma}\big],
\end{align}
such that $\phi_0^{(2)}=\Sigma_{\bm{s}}^{\rho\sigma}X_{\rho\sigma}$. Then Eq.(\ref{delta2}) can be cast into
\begin{align}
\label{delta21}
&\quad\quad\quad\,\;\delta S^{\lambda,\mu\nu}=\frac{1}{2}\int \,d\Gamma \,p^\lambda \Sigma^{\mu\nu}_{\bm{s}}\Sigma_{\bm{s}}^{\rho\sigma}\exp[\xi-\beta\cdot p]X_{\rho\sigma}.
\end{align}
By using Eq.(\ref{f4}), we obtain
\begin{align}
\label{delta22}
&\delta S^{\lambda,\mu\nu}=\frac{1}{m^2}\int \,dP \,p^\lambda (g^{\mu\rho}g^{\nu\sigma}p^2+g^{\mu\sigma}p^{\nu}p^{\rho}+g^{\nu\rho}p^{\mu}p^{\sigma}-[\mu\leftrightarrow\nu])\nn\\
&\quad\quad\;\;\;\times\exp[\xi-\beta\cdot p]X_{\rho\sigma}.
\end{align}

Because integral equation Eq.(\ref{o1}) is not worked out with transition rate unspecified, $\delta S^{\lambda,\mu\nu}$ is expressed as a formal solution to be further determined given specific interaction without losing generality. Qualitatively, one can see it clearly that various thermodynamic forces are responsible for the corrections of spin tensor. Though with rather different tensor structures, involved thermodynamic forces resulting in fluctuation of spin tensor are the same as that in \cite{Bhadury:2020cop}, which can be divided into two groups: one is the group consisting of $\theta,\nabla_\alpha \xi,$ and $\sigma^{\alpha\beta}$, and all members in this group appear in ordinary hydrodynamics. However, their contributions are all proportional to $\Omega$, which reveals that these thermodynamic forces can only affect spin evolution via coupling to spin potential somehow. The other is the group of gradients of spin potential, which is not astonishing as $\Omega$ itself is a Lagranian multiplier for total angular momentum. In the research on spin hydrodynamics, $\Omega$ is always conjugated to spin density and thought to control the evolution of spin.

 \section{Summary and outlook}
 \label{su}
 In this paper, we present a detailed derivation for relativistic first-order spin hydrodynamics using the Chapman-Enskog method to linearize the nonlocal collision term for massive fermions proposed in \cite{Weickgenannt:2021cuo}. This collision term derived from Wigner formalism can provide a natural description of spin-orbit coupling in the collision process and thus is relevant for the research on local spin polarization. With the interaction between fermions unspecified, we give a formal discussion about the motion equations and nonequilibrium corrections to the tensors we are concerned about. Besides the motion equations for basic variables $\mu(x), T(x)$, and $u^\mu(x)$, the motion equation for the newly introduced variable spin potential $\Omega^{\mu\nu}$ is also determined. We find that  the motion equations show no differences compared to spinless first-order hydrodynamics, meanwhile, the energy-momentum tensor receives no corrections from spin and retains the symmetric form as far as first-order theory is concerned. Such results indicate that we need to go over first order to derive spin hydrodynamics, otherwise the effect of spin-orbit coupling would not play the role because the anti-symmetry of energy momentum tensor arises from nonlocal effects in the order of $O(\partial^2)$. It might help that to construct Burnett equations by keeping the expansion to the second order in gradients when $T^{[\mu\nu]}$ enters power counting naturally. However, there are no signs that the acausal problem would be overcome by the new terms introduced by spin. Actually, we obtain the same equations in spinless sector and therefore are also plagued by acausality. So it makes sense that we should turn to  moment method for constructing the second-order theory, which successfully fixes the acausal problem and provides numerically stable hydrodynamic equations. We comment that the deviation function we obtain can be well used to hint the trial function or proper parametrized form for  moment method. The construction of the second-order theory based on our present work and subsequent evaluation of relevant transport coefficients will be performed in  future. There is also one thing that needs to be handled with caution. Generally speaking, most of the interactions we met in quantum field theory is transferred by gauge bosons, which means the gauge link must be plugged into the definition of the Wigner function  to complete the derivation of the collision term.

\section*{Acknowledgments}
J.H. is grateful to Jiaxing Zhao and Ziyue Wang for reading the script and helpful advice.  This work was supported by the NSFC Grant No.11890710, No.11890712 and No.12035006.
\clearpage
\begin{appendix}
	
\section{Thermodynamic  Integral} \label{int}
Thermodynamic integrals we met in this paper are given by 
\begin{eqnarray}
\label{Inq}
I_{nq}(\beta) &\equiv& \frac{2}{(2q+1)!!} \int \frac{\rm dP}{(2\pi)^3}\,(u\cdot p)^{n-2q} (\Delta_{\alpha\beta} p^{\alpha} p^{\beta})^q e^{-\beta \cdot p},
\end{eqnarray}
and noting that $K_n(z)$ denotes the modified Bessel functions of the second kind defined as
\begin{eqnarray}
K_n(z) \equiv \int_0^{\infty} \mathrm{d}x\, \cosh(nx)\, e^{- z \cosh x},
\end{eqnarray}
many thermodynamic integrals $I_{nq}$ can be worked out analytically in the form of $K_n$. 
Specially, we note that $I_{10}=n_0(T)$ , $I_{20}=e_0(T)$, $I_{21}=-P_0(T)$, $I_{30}=T\big(3h_0(T)+z^2P_0(T)\,\big)$ and $I_{31}=-h_0(T)T$ with $h_0(T)\equiv e_0(T)+P_0(T)$.

We also define another thermodynamic integral which is obtained via the following expression
\begin{align}
\label{dnq}
D_{nq}(\xi,\beta)\equiv \exp(\xi)\big(I_{n+1,q}(\beta)I_{n-1,q}(\beta)-I^2_{nq}(\beta)\,\big),
\end{align}
where $D_{nq}$ acts as a Jaccobi determinant in variable transformation. In transforming $Dn$, $De$ into $D\xi$ and $D\beta$, Eq.(\ref{dnq}) is utilized combined with
\begin{align}
\label{pb}
\frac{\partial I_{nq}(\beta)}{\partial \beta}=-I_{n+1,q}(\beta),
\end{align}
which follows from integrating Eq.(\ref{Inq}) by parts.

When handling the thermodynamic integrals with various indices,  the following recurrence relations are very useful to avoid repetitive calculations
\begin{eqnarray}
I_{n,q}&=& \frac{1}{\beta} \left[(n - 2 q) I_{n-1,q} - I_{n-1, q-1}\right],\label{Inqd}\\
DI_{n,q} &=& -  I_{n+1, q} D\beta\,,\label{DI}
\end{eqnarray}	
where Eq.(\ref{DI}) follows directly from Eq.(\ref{pb}).
\clearpage
\section{Calculation of $D\Omega^{\mu\nu}$} \label{do}
When dealing  with the motion equation of spin potential,  we have met a lengthy expression. Here we present 
the different coefficients appearing in Eq.(\ref{Do})
\begin{eqnarray}
R_{\pi}^{[\mu}{}_{\lambda} &=& - \Omega^{[\mu}{}_{\lambda} R_{\pi 1} - u^{[\mu} u^{\alpha} \Omega_{\alpha \lambda} R_{\pi 2},\label{r1}\\
R_{\Pi}^{\mu \nu} &=& R_{\Pi 1} \Omega^{\mu \nu} + R_{\Pi 2} u^{\alpha} u^{[\mu} \Omega^{\nu]}{}_{\alpha},\label{r2}\\
R_n^{[\mu \nu]}{}_{\alpha} &=& - R_{n1}u^{[\mu} \Omega^{\nu]}{}_{\alpha} -R_{n2} g^{[\mu}{}_{\alpha} u^{\kappa} \Omega^{\nu]}{}_{\kappa},\label{r3}\\
R_{\Sigma 1}^{\alpha} &=& - u^{\alpha} \frac{2\, I_{31}}{(m^2\, I_{10} - 2\, I_{31})}, \label{r4}\\
R_{\Sigma 2}^{[\mu \nu] \alpha} &=&  -u^{[\mu} g^{\nu]\alpha}R_{\omega },\label{r5}
\end{eqnarray}
with
\begin{eqnarray}
R_{\pi 1} &=& \frac{4 I_{31}}{(m^2\, I_{10} - 2 I_{31})},\nn\\
R_{\pi 2} &=& \frac{4 (I_{30} - I_{31}) I_{31}}{(m^2\, I_{10} - 2\, I_{31}) \big[m^2\, I_{10} - (I_{30} + I_{31})\big]},\nn\\
R_{\Pi 1} &=& - \frac{1}{\left(I_{10} - \frac{2}{m^2} I_{31}\right)} \left(\frac{I_{20}h-I_{30}n}{D_{20}}  I_{10} - \frac{I_{10}h-I_{20}n}{D_{20}} I_{20} + I_{10} - \frac{2}{m^2} \frac{I_{20}h-I_{30}n}{D_{20}} \, I_{31} + \frac{2\,(I_{10}h-I_{20}n)\, I_{41}}{m^2D_{20}} \right.\nn\\
&&\left.- \frac{10 I_{31}}{3\, m^2}\right),\label{rr1}\\
R_{\Pi 2} &=& \frac{2}{m^2\, I_{10} - 2\, I_{31}} \Bigg[\frac{I_{10}h-I_{20}n}{D_{20}} \!\left(I_{40} - I_{41}\right) - \frac{I_{20}h-I_{30}n}{D_{20}}\! \left(I_{30} - I_{31}\right)  - \left(I_{30} - \frac{11}{3} I_{31}\right) \nn\\
&&+ \frac{\left(I_{30} - I_{31}\right)}{m^2\, I_{10} - I_{30} - I_{31}}\, \times\bigg(\!m^2\, \frac{I_{20}h-I_{30}n}{D_{20}}  I_{10} - m^2 \frac{I_{10}h-I_{20}n}{D_{20}}\, I_{20} + m^2 I_{10} - \frac{I_{20}h-I_{30}n}{D_{20}}\! \left(\!I_{30} + I_{31}\!\right)\!  \nn\\
&&+ \frac{I_{10}h-I_{20}n}{D_{20}}\! \left(\!I_{40} + I_{41}\!\right)\! +\beta I_{41} - \frac{5}{3} I_{31}\bigg)\! \Bigg],\label{rr2}\\
R_{n1} &=& \frac{2}{\left(m^2\, I_{10} - 2\, I_{31}\right)} \left(I_{31} - \frac{n_0\, I_{41}}{I_{20}-I_{21}}\right)+\frac{1}{m^2I_{10} - \left(I_{30} + I_{31}\right)} \left(I_{31} - \frac{n_0 I_{41}}{I_{20}-I_{21}}\right) \frac{2 \left(I_{30} - I_{31}\right)}{\left(m^2\, I_{10} - 2\, I_{31}\right)}, \label{rr3}\\
R_{n2} &=&  \frac{1}{m^2I_{10} - \left(I_{30} + I_{31}\right)} \left(I_{31} - \frac{n_0 I_{41}}{I_{20}-I_{21}}\right) \frac{2 \left(I_{30} - I_{31}\right)}{\left(m^2\, I_{10} - 2\, I_{31}\right)},\label{rr4}\\
R_{\omega } &=&  \frac{2\, (I_{30} - I_{31}) I_{31}}{\left( m^2\, I_{10} - 2\, I_{31}\right) \left[m^2\, I_{10} - \left( I_{30} + I_{31}\right)\right]}+\frac{2\, I_{31}}{ \left(m^2\, I_{10} - 2\, I_{31}\right)} . \label{rr5}
\end{eqnarray}
\clearpage

\end{appendix}
\bibliographystyle{apsrev}
\bibliography{spinhydro}{}

\end{document}